\documentclass[12pt]{article} 
\usepackage[sectionbib]{natbib}
\usepackage{array,epsfig,fancyheadings}
\usepackage[]{hyperref}  
\usepackage{bm,float}
\usepackage{graphicx}
\usepackage{algorithm,comment}
\usepackage[noend]{algpseudocode}
\usepackage[figuresright]{rotating}
\usepackage{sectsty, secdot}
\renewcommand{\theequation}{\thesection\arabic{equation}}

\usepackage{amsmath}
\usepackage{amssymb}
\usepackage{amsfonts}
\usepackage{multirow}
\usepackage{amsthm}

\newtheorem{theorem}{Theorem}

\theoremstyle{definition}

\usepackage[left=3cm, right=3cm, top=2cm]{geometry}

\begin{document}


\renewcommand{\baselinestretch}{2}
\newcommand\T{\rule{0pt}{2.6ex}}       
\newcommand\B{\rule[-1.2ex]{0pt}{0pt}} 

\markright{ \hbox{\footnotesize\rm Statistica Sinica
}\hfill\\[-13pt]
\hbox{\footnotesize\rm
}\hfill }

\markboth{\hfill{\footnotesize\rm FIRSTNAME1 LASTNAME1 AND FIRSTNAME2 LASTNAME2} \hfill}
{\hfill {\footnotesize\rm FILL IN A SHORT RUNNING TITLE} \hfill}

\renewcommand{\thefootnote}{}
$\ $\par


\centerline{\large\bf  Evolutionary State-Space Model and Its Application to}
\vspace{2pt} \centerline{\large\bf Time-Frequency Analysis of Local Field Potentials}
\vspace{.4cm} \centerline{Xu Gao$^{1}$, Weining Shen$^{1}$, Babak Shahbaba$^{1}$, Norbert J. Fortin$^{2}$, Hernando Ombao$^{3}$} \vspace{.4cm} \centerline{\it
$^{1}$Department of Statistics, University of California, Irvine, California, U.S.A.
}  \centerline{\it $^{2}$Department of Neurobiology and Behavior, University of California Irvine, Irvine, California, U.S.A.}
\centerline{\it $^3$Statistics Program, King Abdullah University of Science and Technology, Saudi Arabia.}
\vspace{.55cm} \fontsize{9}{11.5pt plus.8pt minus
.6pt}\selectfont


\begin{quotation}
\noindent {\it Abstract:}
We propose an evolutionary state space model (E-SSM)  
for analyzing high dimensional brain signals whose statistical properties 
evolve over the course of a non-spatial memory experiment. 
Under E-SSM, brain signals are modeled as mixtures of components (e.g., AR(2) process)
with oscillatory activity at pre-defined frequency bands. To account for the potential non-stationarity of these components (since the brain responses could vary throughout the entire experiment), the 
parameters are allowed to vary over epochs. Compared with classical approaches such as independent component analysis and filtering, the proposed
method accounts for the 
entire temporal correlation of the components and  
accommodates non-stationarity. For inference purpose, we propose a novel computational algorithm based upon using Kalman smoother, maximum likelihood and blocked 
resampling. The E-SSM model is applied to simulation studies and an application to a multi-epoch local field potentials (LFP) signal data collected from a non-spatial (olfactory) sequence memory task study. The results confirm that our method captures the evolution of the power for different 
components across different phases in the experiment and identifies clusters of electrodes that behave 
similarly with respect to the decomposition of different sources. 
These findings suggest that the activity of different electrodes does
change over the course of an experiment in practice; treating these 
epoch recordings as realizations of an identical process could lead to misleading results. In summary, the proposed method underscores the importance of capturing the evolution in brain responses over the study period.

\vspace{9pt}
\par
\end{quotation}\par

\def\thefigure{\arabic{figure}}
\def\thetable{\arabic{table}}

\renewcommand{\theequation}{\thesection.\arabic{equation}}


\section{Introduction}
\label{introduction}
The goal of this paper is to develop a novel statistical model for investigating the evolution of a brain process duration of a learning experiment. 
To infer brain neuronal activity, electrophysiological recordings such as local field potentials (LFPs) and electroencephalograms 
(EEGs) are commonly used to indirectly measure electrical activity of neurons. In this paper, we consider LFPs from multiple electrodes that capture the integration of membrane currents in a local region of cortex \citep{mitzdorf1985current}. 

\begin{figure}[h] 
	\centering
	\includegraphics[width=.7\textwidth,height=0.3\textheight]{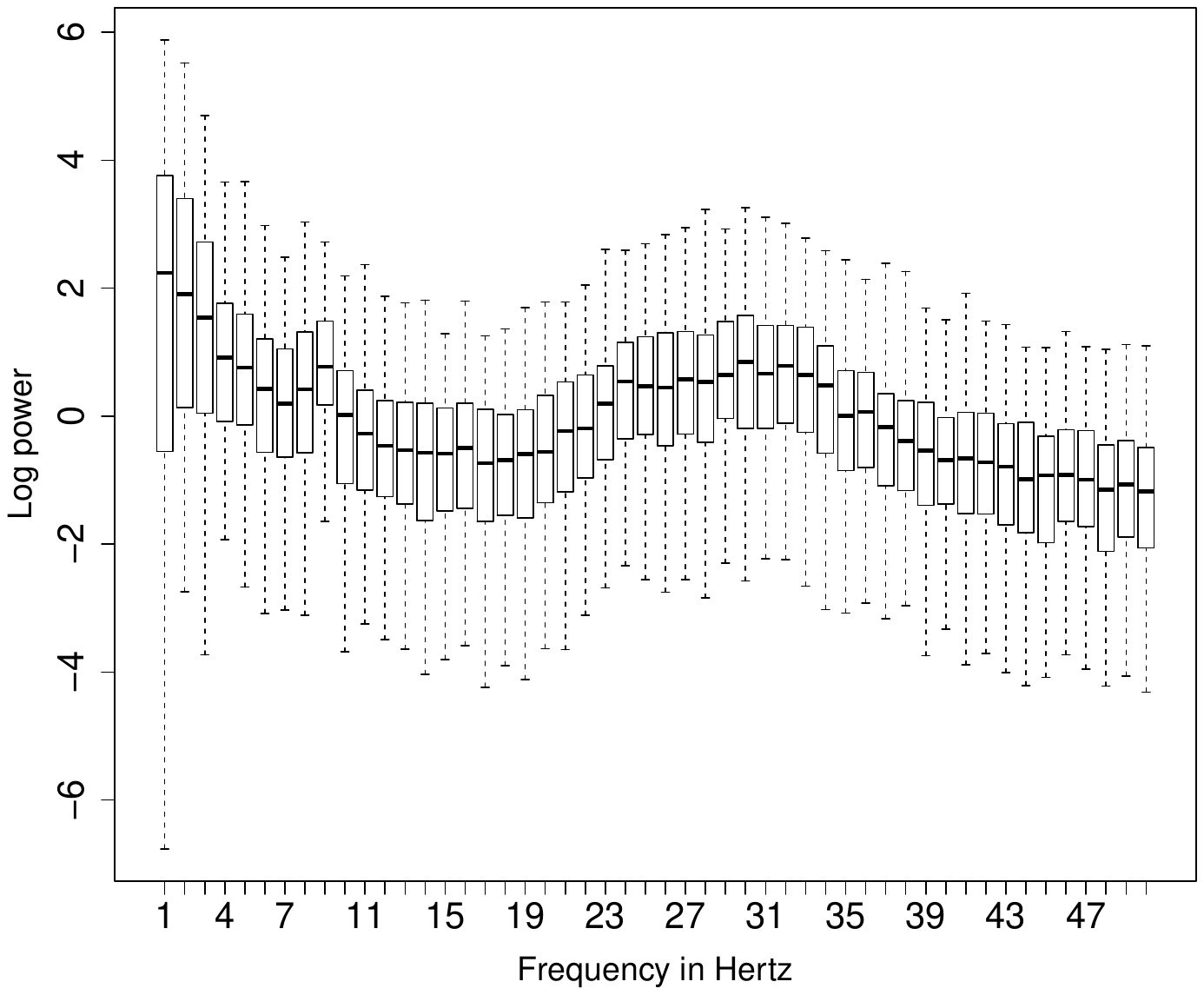}
	\caption{The log periodogram boxplots for each frequency obtained by all 247 epochs at electrode T22.}
	\label{Intro_ex2}
\end{figure}

\begin{figure}[h]
	\centering
	\begin{tabular}{cc}
		\includegraphics[scale=0.25]{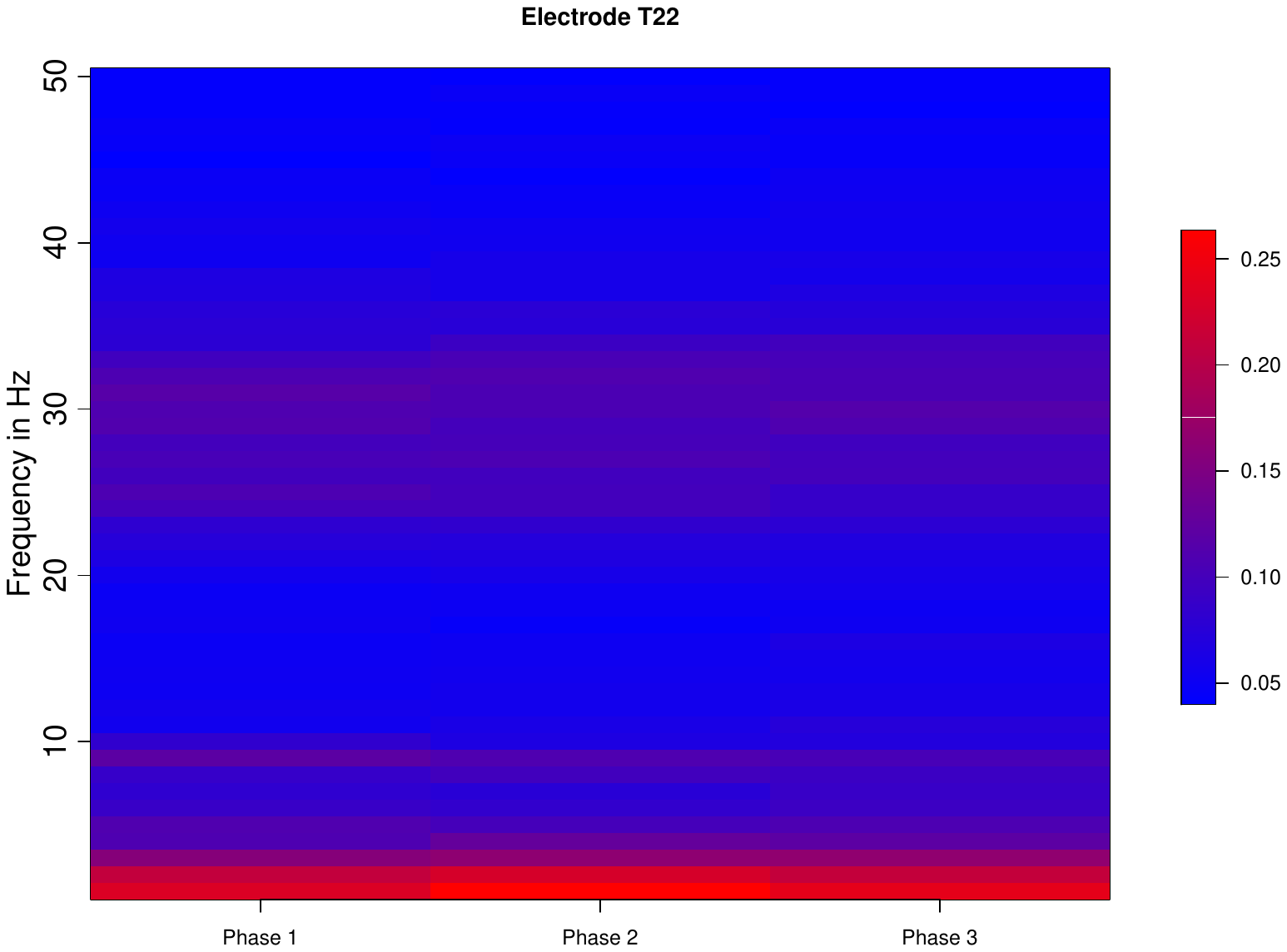}&\includegraphics[scale=0.25]{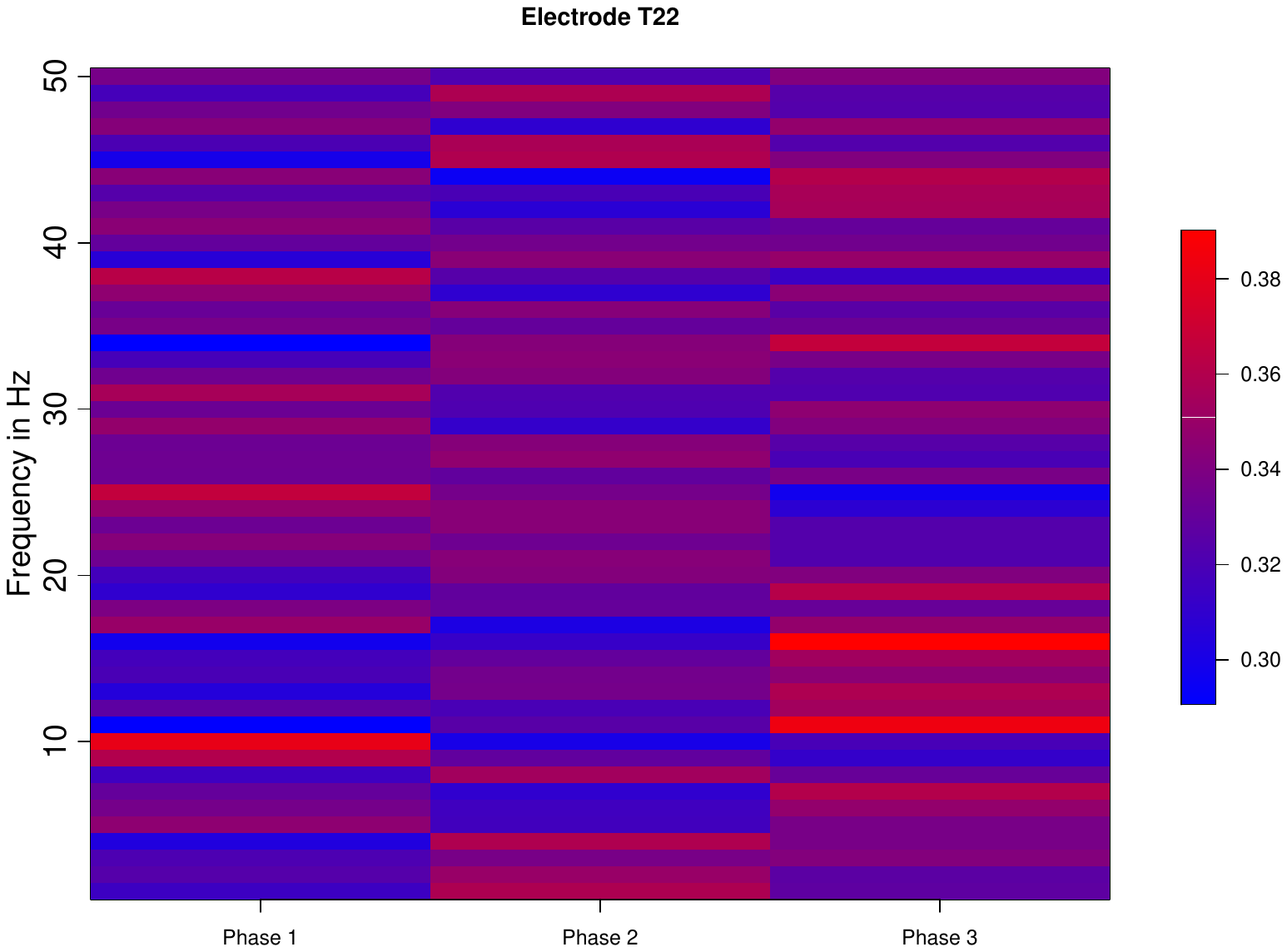}\\
	\end{tabular}
	\caption{Left: The heatmap of the averaged periodogram among Phase 1 (epochs 1 - 80), Phase 2 (81 - 160) and Phase 3 (161 - 247) respectively at electrode T22. The original signals were rescaled to unit variance. Right: The heatmap of the relative periodogram (summing up to 1 for each frequency).  Spectral power (decomposition of waveform) evolved across phases of the experiment.}
	\label{Intro_ex3}
\end{figure}

In practice, LFPs are the observed spatio-temporal signals at different tetrodes. In a motivating example, an olfactory (non-spatial) sequence memory experiment has been performed in a memory laboratory to study how neurons learn the sequential ordering of presented odors \citep{allen2016nonspatial}. In this study, LFP recordings in a rat are obtained from an implanted plate with 12 electrodes. One epoch corresponds to about 1 second in physical time. We further study the behavior of these LFPs by examining their spectra. In Figure~\ref{Intro_ex2}, we plot the boxplots of the log periodograms across all the epochs from one electrode. These plots reveal that LFPs contain power at distinct bands: delta (0-4 Hertz), alpha (8-12 Hertz) and the high-beta low-gamma (30-35 Hertz) bands. As an exploratory step, we divide the entire experiment into three phases, early, middle, and late phases. In each phase, we compute the average periodogram (averaged across epochs) and present them on the left side of Figure~\ref{Intro_ex3}. On the right side, we plot the relative periodogram (obtained by rescaling the periodogram so that the relative periodogram for each frequency sums up to 1) and find that the spectral power evolves during the course of experiment. During the early phase, power has a broad (rather than concentrated) spread across bands. However, at the late phase, power seems to be more concentrated at the lower beta band.

In summary, preliminary results suggest that there exists a strong similarity of the LFP waveforms across many electrodes. Moreover, the spectra of the LFPs appear to change across the epochs in the experiment. In a recent study, \cite{gao2018regularized} proposed a matrix data clustering approach and the results also indicated the existence of spectra heterogeneity. Therefore, statistical models that are capable of describing LFP signals' evolution over the course of epochs are largely needed to help understand how the rat learns the sequence of the odor presentation. 

In the literature, LFPs and other electrophysiological signals are commonly characterized as mixtures of different underlying brain oscillatory processes and there have been a number of 
approaches used to estimate these latent independent sources  
\citep{whitmore2016unmasking, einevoll2007laminar,prado2013sequential}. For example, data-adaptive methods such as independent components analysis 
(ICA) and principal components analysis (PCA) can provide estimates for the unobserved cortical sources. However, they usually do not take into account the spectral structure within underlying sources that could evolve over the course of the experiment given multiple epochs. Moreover, 
without any constraint on the structure of the sources, it is 
extremely difficult to pool information across the epochs in the 
experiment. Recently, \cite{fiecas2016modeling} studied the dynamics of LFPs 
during the course of experiment via Cram\'er representations. Their approach does not consider low-dimensional representations, which 
are indispensable to modeling these high dimensional multi-electrode LFPs. 

To overcome the aforementioned limitations, we develop an evolutionary state space model (E-SSM) that explicitly captures the evolutionary behavior in high dimensional time series. The E-SSM shares a similar form with the classical state-space model (as in \citet{shumway2013time}) but differs in that the parameters are varying across epochs and the mixing matrix is unknown and therefore has to be estimated.  Moreover, E-SSM manages to capture the temporal correlation of each of the latent sources by characterizing them using second order autoregressive [AR$(2)$] processes. The reason for choosing AR$(2)$ is due to its ability to capture the precise oscillatory behavior of these latent sources. In particular, by parameterizing these sources as AR$(2)$, we can easily constrain the power of each source to center at pre-specified frequency bands such as delta (0 - 4 Hertz), alpha (8 - 12 Hertz) and high-beta gamma ($>$ 30 Hertz) bands, where the choice of these particular frequency bands is due to the standard convention in neuroscience based upon previous electrophysiological data analysis \citep{deuschl1999recommendations}. The use of AR$(2)$ mixture here can be viewed as an analogy of Gaussian mixture models for classical density estimation problems. Compared to the classical methods such as ICA and PCA, the sources produced by E-SSM are more directly interpretable in terms of oscillatory properties. 

The main contributions of this paper are as follows: (1) The proposed 
E-SSM model provides a rigorous framework in modeling brain activity, connectivity and their dynamic behavior during the course of experiments. 
In particular, our model accounts for the temporal evolution/dependence 
of the spectrum power for particular frequency bands across the 
entire experiment as well as the temporal structure among 
the latent sources. (2) E-SSM gives interpretable results by modeling particular 
predominant frequency bands that are associated with various brain functional states through AR$(2)$ processes. (3) In theory, we show that the spectrum of arbitrary weakly stationary time series can be approximated by the spectrum of AR(2) mixtures, which gives a theoretical justification of the use of AR(2) mixtures. We also give a strong consistency result for the MLE of E-SSM. (4) By applying the E-SSM model, one can easily conduct analysis on both of time 
and frequency domains and thus provide a complete characterization of the underlying brain process. (5) Finally, the E-SSM model and the 
proposed estimation method, in general, are intuitive and can be 
implemented easily thanks to the existing theory and algorithm 
for state space model. However, the key difference is the generalization of the
multiple epochs setting which allows pooling information across 
epochs and a flexible mixing matrix estimation step. 



\section{Evolutionary State Space Model (E-SSM)}\label{section_ARG}
In this section, we discuss models for inferring latent structures in LFPs and their evolution across epochs over the entire experiment. We shall first describe the model for a single epoch and then discuss the extension to treat multiple epochs. 

\subsection{State Space Model for a single epoch}
\label{section_single}
Denote $t = 1, \cdots, T$ as the time points in a single-epoch and \\
$ \bm{Y}_{t} = (Y_{t}(1), \cdots, Y_{t}(p))'$ as the observed LFPs where $p$ is the 
number of electrodes. For any fixed time point $t$, we assume that $\bm{Y}_{t}$ is a {\it mixture} 
of $q$ latent independent source signals $\bm{S}_t = ({S}_t(1), \cdots, 
{S}_t(q))'$, where $q$ is the number of spatial source signals. We assume $p \geq q$. Then the model can be presented as
$
\bm{Y}_{t} = M \bm{S}_t + \bm{\epsilon}_t, 
$
where $M$ is the mixing matrix, $\bm{\epsilon}_t = (\epsilon_t(1), \cdots, \epsilon_t(p))'$ is noise that follows $N(\bm{0}, \tau^2\bm{I_p})$ and $\bm{I_p}$ is an identity matrix of dimension $p$. Each of the independent latent signals ${S}_t(l), l=1, \cdots, q$ models the 
source that represents oscillatory activity at a set of pre-specified frequency bands (e.g., delta, alpha and gamma).

\noindent \underline{Modeling the source signals $\bm{S}_t $} 

One important parameterization in our model is to constrain the sources to have an AR(2) structure such that each represents a particular oscillator: delta $(\delta$: 0 - 4 Hertz), theta $(\theta$: 4 - 8 Hertz), alpha $(\alpha$: 8 - 12 Hertz), lower beta $(\beta$: 12 - 18 Hertz) and gamma $(\gamma$: $>30$ Hertz). Recall that an autoregressive operator of order 2 is defined by
\begin{equation}
\label{poly}
\phi(B) = 1 - \phi_1B - \phi_2 B^2,
\end{equation}
where $B$ is a backshift operator defined by $B^{\ell} S_t = S_{t-\ell}$, and $\phi_1,\phi_2$ are the corresponding coefficients. It can be shown that the spectrum of an  AR(2) process with noise level $\sigma_{\omega}$ is $f_S(\omega) = \frac{\sigma_w^2}{|1- \phi_1\exp(-2\pi i \omega) - \phi_2 \exp(-4\pi i \omega)|^2}.$ To illustrate its use in practice, we plot the spectrum of an  AR(2) process 
with $\phi_1 = 1.976, \phi_2 = -0.980, \sigma_w = 0.1$ in Web Figure~20. It can be seen that there is a peak 
at frequency $\omega = 10$ Hertz, which means
that the frequency band around $\omega = 10$ Hertz dominates the process and thus produces the most power. This property of AR$(2)$ model makes 
it potentially useful for characterizing brain signals (such as LFPs) 
with oscillations at either broad or narrow frequency band. 

We now explain the connection between the AR$(2)$ coefficients and the spectrum (i.e., the location and spread of the peak). 
First, the process is causal when the roots of the polynomial in Equation~(\ref{poly}) have magnitudes greater than 1.
Furthermore, under causality, 
\cite{jiru2008relationships} and \cite{shumway2013time} demonstrate 
that when the roots of the polynomial in Equation~(\ref{poly}) are 
complex-valued with magnitude greater than 1, then the 
spectrum attains a peak that is centered around the phase of the roots. Moreover, when the magnitude of the roots 
become larger than $1$, the peak becomes less concentrated around the 
phase.

Motivated by this result, we will fix the phase (or argument) of each  AR(2) polynomial root to model each of the particular bands obtained from previous study results. As noted, fixing the phase is consistent with neuroscience standard and thus will not be a constraint in practice. In the field of neuroscience, neural oscillations are widely captured at all levels such as LFPs, EEG and neuro spike trains \citep{busch2009phase}. To characterize those oscillatory patterns, one typical approach is to convert the original electrophysiological signals to the spectrum domain by Fourier transformation. In this way, oscillations can be represented by modulus and phases. Among all the frequencies, the first interesting frequency band (alpha band) is introduced by \cite{gerrard2007mechanisms}. Later on, other bands including delta, theta, beta, gamma are being studies from various perspectives. As shown in the motivating example in Section 1, our collaborators from neuroscience studied the frequency domain behavior and concluded that ``low-gamma oscillations were more strongly modulated by temporal context and performance than theta oscillations " \citep{allen2016nonspatial}. Thus, by fixing the phase in our model, we are able to maintain consistent with neuroscience standard and thus will not be a constraint in practice.
To model the evolution across epochs, we allow the modulus of the  AR(2) polynomial roots to change among epochs. As a result, as the phase of the roots for each of the latent independent source signals is fixed, the  AR(2) process is uniquely determined by the modulus 
and the variance. In practice, the value of modulus controls the spread of the spectrum curves. 
For an  AR(2) process $S_t = \phi_1S_{t-1} + \phi_2S_{t-2} + w_t$, the modulus $\rho$ 
and phase $\psi$ of the roots of the polynomial have the relationship that 
$\phi_1 = 2\rho^{-1}cos(\psi)$ and $\phi_2 = -\rho^{-2}$. This result can be seen as an analogy of the use of Gaussian mixture model (or any location-scale mixture in general) for density estimation.  

\vspace{.1in}
\noindent \underline{Generalized state-space model} \\
Following the previous discussion, the latent independent spatial source signals are modeled as multivariate  AR(2)s, 
$
\bm{S}_t = \Phi_1 \bm{S}_{t-1} + \Phi_2 \bm{S}_{t-2} + \bm{\eta}_t, 
$
where $\Phi_1 = \text{diag} (\phi_{11}, \cdots, \phi_{q1}), \Phi_2 = \text{diag} (\phi_{12}, \cdots, \phi_{q2}) \in \mathbb{R}^{q\times q}$ are diagonal matrices, and the noise $\bm{\eta}_t = (\eta_1(t), \cdots, \eta_q(t))' \sim N(\bm{0}, \sigma^2\bm{I_q})$. The final model can hence be viewed as a generalized state-space model: 
\begin{align}
\label{single_model}
\begin{split}
&\bm{Y}_{t} = \widetilde{M} \bm{X}_t + \bm{\epsilon}_t, \\
&\bm{X}_t = \widetilde{\Phi} \bm{X}_{t-1} + \widetilde{\bm{\eta}}_t,
\end{split}
\end{align}
where $\bm{X}_t = (\bm{S}'_t , \bm{S}'_{t-1} )',$ $\widetilde{M} = (M, \bm{0}) \in \mathbb{R}^{p*2q},$ $\widetilde{\Phi} = \begin{bmatrix}
\Phi_1 & \Phi_2 \\
\bm{I_q}&\bm{0}
\end{bmatrix},$ and $\widetilde{\bm{\eta}}_t = (\bm{\eta}'_t, \bm{0})'$. Note that the residual $\bm{\epsilon_t}$ is assumed to be independent across time $t$. It implies that all the temporal correlations are characterized by the underlying latent signals $\bm{S}_t$. The model in \eqref{single_model} is not a regular state-space model since the mixing matrix $\widetilde{M}$ is unknown. Moreover, following the aforementioned discussion, the coefficients of the autoregressive processes are determined by the modulus $\bm{\rho} = (\rho_1, \cdots, \rho_q)$ and phase $\bm{\psi} = (\psi_1, \cdots, \psi_q)$ of the autoregressive polynomial roots. Since we are interested in particular frequency bands, we fix the phase $\bm{\psi}$ and the state equation in~(\ref{single_model}) is parameterized by $\bm{\rho}$ and $\sigma^2$.

\subsection{Evolutionary State Space Model for multiple epochs}
\label{section_multiple}
Next, we generalize the model in Section~\ref{section_single} to accommodate multiple epochs. We assume that across epochs, the mixing matrix $M$ is fixed and the latent independent autoregressive processes evolve through the modulus $\bm{\rho}.$ This assumption implies that the cortical structure remains unchanged across epochs for each individual. We denote $r = 1, \cdots, R$ as the epochs in the experiment, then the model is given by
\begin{align}
\label{multiple_model}
\begin{split}
&\bm{Y}^{(r)}_{t} = \widetilde{M} \bm{X}^{(r)}_t + \bm{\epsilon}^{(r)}_t, \\
&\bm{X}^{(r)}_t = \widetilde{\Phi}^{(r)} \bm{X}^{(r)}_{t-1} + \widetilde{\bm{\eta}}^{(r)}_t,
\end{split}
\end{align}
where the definition of $\bm{Y}^{(r)}_{t}, \widetilde{M}, \bm{X}^{(r)}_t, \widetilde{\Phi}^{(r)}, \bm{\epsilon}^{(r)}_t, \widetilde{\bm{\eta}}^{(r)}_t$ are similar as in Equation~(\ref{single_model}) except the additional superscript $r$ for each epoch $r$. 

In the proposed model, we assume an autoregressive structure that evolves across epochs. This assumption is inspired by the preliminary analysis in Section~\ref{introduction} showing that the power spectrum evolves during the course of the experiment. Accordingly, the evolutionary spectrum of each latent source will be easily captured in an explicit form\\ $f^{(r)}(\omega) = \frac{\sigma_w^{2(r)}}{|1- \phi_1^{(r)}\exp(-2\pi i \omega) - \phi_2^{(r)} \exp(-4\pi i \omega)|^2}$. We also assumed that the mixing matrix is invariant to epochs. This is due to the fact that the network structure of subjects is not changing across phases of experiments. To reiterate, non-stationarity  will be captured by the AR(2) coefficients.


In the literature, there have been numerous discussions on the identifiability issues of state-space models \citep{hamilton1994time}. Indeed, for a general state-space model, the same representation can be obtained by applying an orthogonal transformation on matrices. \cite{zhang2011general} proposed a non-Gaussian constraint to avoid the identifiability issue. In this paper, to ensure the uniqueness of the solution, we require that each component of the latent independent source signals $\bm{S}(t)$ to have unit variance and the entries of $\widetilde{M}$ are positive.


\section{Inference for E-SSM}
\label{Estimation}

\subsection{Estimating E-SSM}
For E-SSM with single epoch, we propose an iterative algorithm that comprises of Kalman filter and least squares for parameter estimation purpose. More details are given in Section 3 of the Supplementary File. 

Next we extend the previous method to the multiple epoch setting in Equation~\eqref{multiple_model}. The major challenge lies in pooling information from different epochs in estimating the epoch-invariant mixing matrix. Inspired by the resampling approach used for modeling time series with Gaussian process \citep{gao2017modeling} and linear mixed model \citep{cheng2014exact}, we propose a blocked resampling based approach. The key idea can be summarized as follows: we first divide the epochs into blocks; then for each block we estimate the corresponding mixing matrix and the epoch-specific AR(2) parameters. 
These blocks retain the temporal sequence of the epochs and the final estimate at a previous epoch serves as the initial estimate of 
mixing matrix at the current epoch. 
The final estimates of the mixing matrix obtained from each block 
are averaged to produce the estimate for 
the common mixing matrix. For the next step, given the estimated mixing matrix, we follow Algorithm 1 to obtain estimates of the epoch-specific AR(2) parameters. The  approach is summarized below.

\begin{enumerate}
\item[{\bf II.A}] We fix the length of the blocked resampling sampler as $l$. We draw the starting 
epoch index $s$ from the set $\{1, 2, \cdots, R-l+1\}$. Then at current iteration, the blocked resampling sampler is $(\{\bm{Y}^{(s)}_{t}\}_{t=1}^T, \cdots, \{\bm{Y}^{(s+l-1)}_{t}\}_{t=1}^T)$.

\qquad {\bf A.1}. Starting with epoch $s$, we implement the approach for single epoch in Section~\ref{section_single} on $\{\bm{Y}^{(s)}_{t}\}_{t=1}^T$to obtain estimates $\widetilde{M}^{(s)}$.

\qquad {\bf A.2}. Staring with epoch $s+1$ and the initial value $\widetilde{M}^{(s)}$, we repeat A.1 to obtain estimates $\widetilde{M}^{(s+1)}$.

\qquad {\bf A.3}. We repeat A.2 until the last epoch $s+l-1.$ We denote the final estimates $\widetilde{M}^{(s+l-1)}$ as the ultimate estimates of resampling sampler $(\{\bm{Y}^{(s)}_{t}\}_{t=1}^T, \cdots, \{\bm{Y}^{(s+l-1)}_{t}\}_{t=1}^T).$
The pipeline of the procedure is summarized below.
\begin{equation*}
\begin{bmatrix}
{\bm{Y}}^{(s)}_{1}\\
{\bm{Y}}^{(s)}_{2}\\
\cdots\\
{\bm{Y}}^{(s)}_{T}
\end{bmatrix}
\rightarrow
\widetilde{M}^{(s)}
\rightarrow
\begin{bmatrix}
{\bm{Y}}^{(s+1)}_{1}\\
{\bm{Y}}^{(s+1)}_{2}\\
\cdots\\
{\bm{Y}}^{(s+1)}_{T}
\end{bmatrix} 
\rightarrow\\
{\widetilde{M}^{(s+1)}} 
\cdots
\rightarrow
\begin{bmatrix}
{\bm{Y}}^{(s+l-1)}_{1}\\
{\bm{Y}}^{(s+l-1)}_{2}\\
\cdots\\
{\bm{Y}}^{(s+l-1)}_{T}
\end{bmatrix} 
\rightarrow
\widetilde{M}^{(s+l-1)}
\end{equation*}

\item[ {\bf II.B}.] Repeat II.A until a sufficient number of resampling estimates is obtained. 
Compute the average of those estimates, defined by $\widetilde{M}_{g}$, as the global estimate of 
$\widetilde{M}$.

\item[\noindent {\bf II.C}.]  Plug the global estimate $\widetilde{M}_g$ into every single epoch. 
Following Algorithm 1 for single epoch, we
obtain the estimates of $\bm{\rho}^{(r)}, {\sigma^2}^{(r)}, {\tau^2}^{(r)}, r=1, \cdots, R.$
\end{enumerate}

The over-all work flow is given in Web Figure 10 (Supplementary File). Note that since the 
mixing matrix $\widetilde{M}$ are the same across epochs, we use the blocked resampling 
strategy to get the global estimates sequentially. Given that estimate, we proceed to make 
inference on every single epoch. For the choice of the length $l$, we recommend starting from $l = CR^{1/k}$ and then increasing $l$ until a stable result is obtained, where $k = 3, 4$, $C$ is a constant and $R$ is the number of epochs.


\subsection{Testing for difference across epochs}
Inspired by the preliminary results shown in Figure 3, we assumed that all the epochs can be divided into different 
phases, among which there exist discrepancies in $\widetilde{\Phi}^{(r)}$. In order to test whether such a difference in $\widetilde{\Phi}^{(r)}$ is significant across different
phases, we propose a permutation test by shuffling epochs between phases and then  implementing E-SSM to obtain parameter estimates and their reference distributions. We give a simulation example to demonstrate its use in Section 6.

\section{Theory} 
We start with a strong consistency result for the MLE of the proposed E-SSM model. Denote $\Theta = (M,\widetilde{\Phi}^{(1)},\ldots,\widetilde{\Phi}^{(R)})$ as the collection of parameters in the multiple epoch model \eqref{multiple_model}. Let $\hat{\Theta} $ and $\Theta_0 = (M_0,\Phi_0^{(1)},\ldots,\Phi_0^{(R)}) $ be the MLE and the true value of $\Theta$, respectively. Theorem \ref{thm:mle} below states that under mild conditions, $\hat{\Theta}$ is a strongly consistent estimator for $\Theta_0$. 
\begin{theorem}\label{thm:mle}
 Suppose that the AR(2) process in the definition of $\bm{S}_t^{(k)}$\eqref{multiple_model} is causal for every epoch $k=1,\ldots,R$. Assume $M_0$ is of full column rank, $(\Phi_0^{(1)},\ldots,\Phi_0^{(R)})$ is of full row rank, and every parameter (matrix) in $\Theta_0$ belong to a known compact support. Then $\hat{\Theta}$ converges to $\Theta_0$ almost surely. 
\end{theorem}
The assumptions in Theorem \ref{thm:mle} are easily satisfied for many situations. For example, in our case, $\Phi_0^{(i)}, i=1,\ldots,R$ are diagonal matrices with elements being AR(2) coefficients centered at pre-specified frequency bands. As long as these bands are different, this assumption is satisfied.  The consistency result also applies for single epoch model \eqref{single_model} by letting the number of epochs $R=1$. The proof of the theorem, which we defer to the Supplementary File, is based on the consistency results for general hidden Markov model in \citet{Douc2011}. 
Next we give an AR(2) decomposition theorem stating that the spectrum of any weakly stationary process can be approximated by that of a linear mixture of AR(2) processes. 
This result provides a theoretical justification for representing individual sources by AR(2) models due to their ability to present each source signal at pre-specified frequency bands. 
\begin{theorem}\label{thm1}
	Let $Y_t$ be a weakly stationary time series with zero mean and continuous spectrum $f_Y(\omega).$ Let $0=\omega_0 < \omega_1 < \cdots < \omega_J = 1/2$, and $\xi = \max_{j=1}^J | \omega_j - \omega_{j-1}|$. 
	Denote $S_t^{(j)}, j=1, \cdots, J$ as independent AR(2) processes with unit variance and spectrum  of $f_{S^{(j)}}(\omega)$  such that the phase of its AR polynomial roots, denoted by $\psi^{(j)}$, satisfies $\psi^{(j)} \in [\omega_{j-1}, \omega_j)$. Consider a family of processes $\{Q_{t,J}\}_{J=1}^\infty$ defined by $Q_{t,J} = \sum_{j=1}^{J} {a}_j S_t^{(j)}$ with non-negative coefficients $\{a_j\}_{j=1}^J$ and let $\mathcal{F}_J$ be the collection of spectrum of $\{Q_{t,J}\}$. Assume that $\xi \rightarrow 0$ as $J \rightarrow \infty$, then  
	\begin{equation}
	\inf_{f \in \mathcal{F}_J} \|f_Y  - f \|_{2}  \rightarrow 0~~~\text{as}~J \rightarrow \infty.
	\end{equation} 
Moreover, if $f_Y$ is Lipschitz continuous, and $\omega_k = k/(2J)$ for $k=0,\ldots,J$. Then for any sufficiently large $J$ and some positive constant $C$,	\begin{equation}
	\inf_{f \in \mathcal{F}_J} \|f_Y  - f \|_{\infty}  < C J^{-1}.
	\end{equation}  
\end{theorem}
Theorem \ref{thm1} states that the minimum approximation error of the spectrum from a class of finite mixture AR(2) models is negligible given the number of terms $J$ goes to infinity. In other words, the AR(2) mixture gives a consistent estimate for the spectral density given that $J$ is chosen sufficiently large. Moreover, if we assume that the frequencies $\omega_k$'s are equally spaced, then the convergence rate is essentially equivalent with that of the equally-spaced Fourier series based on Jackson-type of inequality. The convergence rate for finite Fourier series with non-uniformly spaced frequency bands is still unknown to the best of our knowledge \citep{Ep2005}. The proof of theorem is given in Section 2 of the Supplementary File. 

\section{A Comparison to Existing Methods}\label{Connection}
We discuss a few major differences between our method and the existing state-of-art approaches including ICA and classical state-space models. 

ICA has been widely used in single/between-subject electrophysiological exploratory analysis. For example,  \cite{makarova2011parallel} proposed an ICA method to segregate pathways with partially
overlapped synaptic territories from hippocampal LFPs. To investigate the variability across different subjects or subgroups, \cite{guo2011general} proposed a general group probabilistic ICA (pICA) framework with its extensions (\citealp{wang2018hierarchical,lukemire2018hint}) to accommodate cross-subject structure in multi-subject spatial-temporal brain signals. 
Although these methods work well under certain settings, there is still plenty of room for improvement in modeling electrophysiological signals. First, they do not have a mechanism for capturing how the parameters 
(and spectral properties) of the latent source signals evolve across epochs over 
the entire experiment. Most of the existing methods are based on concatenating the signals 
from different epochs and estimating parameters as though these signals 
are realizations of the {\it same} underlying process. However, since the ``reconstructed''
latent sources vary across epochs, there is no rigorous framework for modeling how these 
parameters could change across epochs. 
Second, existing methods do not take into account 
the temporal structure of the latent sources. In fact, these sources are estimated for each 
time point independently of  other time points. Third, current 
ICA methods for source modeling may not produce interpretable results from spectral analysis of electrophysiological signals.
In fact, brain researchers have observed association between power at 
different frequency bands and brain functional states \citep{michel1992localization}. 
Thus, it is necessary to develop a framework that accounts for the evolution of the power 
at these frequency bands over many epochs. Lastly, there are limitations in the connection 
between time and frequency domain analysis. Methods from time and frequency domain are 
developed almost exclusively from each other, which is counter-intuitive since these 
two approaches ought to be used concurrently in order to give a complete characterization 
of brain processes.

\section{Simulation Studies}
\label{section_sim}
\subsection{Results on single epoch analysis}
\label{section_sim_single}
We first evaluate the proposed E-SSM under single epoch setting. We simulate data from three independent  AR(2) processes that corresponds to delta $(\delta$: 0 - 4 Hertz), alpha $(\alpha$: 8 - 12 Hertz), and lower beta $(\beta$: 12 - 18 Hertz). We randomly generate a positive ``mixing'' matrix $M$ and fix the number of electrodes of the observational brain signals to be $20$. In summary, following the notation in Section~\ref{section_single}, we have: $p = 20, T = 1000, q = 3, \tau^2 = 1, \sigma^2 = .1, (\rho_1, \psi_1) = (1.0012, 2), (\rho_2, \psi_2) = (1.0012, 8), (\rho_3, \psi_3) = (1.0012, 15).$

We implement the proposed method in Section~\ref{section_single} and plot the periodograms of the true and reconstructed signals in Web Figure 1 (supplementary file). 
The estimated source signals share exactly the same shape as the true signals. We also compare the results with those of ICA in Web Figure 11. It is clear that ICA is unable to recover the three latent bands while our method manages to separate different spectral components very accurately. These results are consistent with our discussion on the possible drawbacks of ICA.

\subsection{Results on multiple epoch analysis}
\label{sim_multiple}
We then evaluate the performance of the proposed method for multiple epochs. We choose $20$ electrodes and $3$  latent independent  AR(2) processes. To model the evolution across epochs, we allow the modulus $(\rho_1^{(r)}, \rho_2^{(r)}, \rho_3^{(r)})$ increase from $(1.001, 1.001, 1.001)$ with an increment of $0.00005$ as the epoch $r$ propagates. All the remaining parameters are the same as in Section~\ref{section_sim_single}. 
Web Figure 2 (Supplementary File) shows the heatmap of periodogram from electrode $1$ as epochs evolve. The results look satisfactory. 
Web Figure 3 shows the periodograms of the true and estimated signals from the three underlying  AR(2) processes. For the delta, alpha, and lower beta bands, we can see the peaks at the corresponding dominating frequency from the true and estimated signals. As the epochs evolve, we find that both the true and estimated periodograms spread out around the dominating frequency, which indicates that the pattern of the periodograms from the reconstructed  AR(2) process is consistent with that of the true  AR(2) process.  

We also applied ICA to the simulated dataset and presented the results in Web Figures 12 and 13. As expected, ICA hardly separates the three underlying latent sources and rarely captures the spread of power as epoch evolves. This phenomenon coincides with our previous discussion that ICA neglects the dynamics across epochs.

\subsection{Results for settings derived from the data}
Here we simulate the data using parameter setting from the motivating sequence memory study example. We use the estimated modulus \\$(\hat{\rho}_1^{(r)}, \hat{\rho}_2^{(r)}, \hat{\rho}_3^{(r)})$, variances $(\hat{\sigma}^{2(r)}, \hat{\tau}^{2(r)})$ and mixing matrix $\widetilde{M}$ to generate signals across $12$ electrodes among $247$ epochs. To evaluate the performance of E-SSM, we also apply the classical state space model (SSM) estimation methods as a benchmark in comparison with E-SSM. Specifically, we fit SSM for each single epoch and obtain the epoch-specific  parameter estimates. Note that this is the approach that most of the existing methods will use when analyzing signals with multiple epochs. As an alternative, we also compute the average of epoch-specific estimates. 

We compare mean of sum of square errors (MSE) of the parameters
obtained from E-SSM and SSM. In Table~\ref{sim_table}, it is clear that E-SSM successfully captures the evolution of parameters compared to classical state space models. Among all the frequency bands, the benefits are dramatic. These results highlight the advantages of using E-SSM when signals are comprised of multiple epochs. Meanwhile, it also indicates the potential loss of information if we naively average over all the epochs when conducting analysis. 

As a comparison, we also applied ICA to the simulated dataset. The results in Web Figures~14 and 15 suggest that our method manages to estimate the mixing matrix very accurately, while ICA misses most of the patterns across electrodes. Web Figures~16 and 17 show the periodograms obtained from our method and ICA. Again, ICA is unable to recover the true signals or identify the dynamics across epochs.

\begin{table}[H]
	\centering
	\caption{MSE obtained from E-SSM and SSM (benchmark)}
	\label{sim_table}
	\begin{tabular}{llll}
		\hline \T\B
		Parameters	& E-SSM & SSM (average) & SSM (single)  \T \B \\
		\hline
		$\widetilde{\Phi}$ (delta band)	& $\bm{3.33 \times 10^{-5}}$ & $7.27 \times 10^{-5}$ & $5.53 \times 10^{-5}$ \T\B \\
		$\widetilde{\Phi}$ (alpha band)	& $\bm{1.41 \times 10^{-5}}$ & $3.23 \times 10^{-5}$  &$2.89 \times 10^{-5}$ \T\B \\
		$\widetilde{\Phi}$ (gamma band)		& $\bm{1.69 \times 10^{-5}}$ &  $8.07 \times 10^{-5}$ & $2.00 \times 10^{-5}$\T\B \\
		$\tau^2$	& $\bm{9.31 \times 10^{-6}}$ & $2.03 \times 10^{-4}$ & $1.91 \times 10^{-4}$ \T\B \\
		$\sigma^2$		& $\bm{1.93 \times 10^{-1}}$ &  $1.93 \times 10^{-1}$ &$1.91 \times 10^{-1}$  \T\B \\
		\hline
	\end{tabular}
\end{table}

\subsection{Results on permutation test}
\label{sim_test}
Following similar strategies in previous simulations, we generated 5 latent AR(2) processes corresponding to delta, theta, alpha, lower bet and gamma bands. We assumed that there were two phases with 40 epochs in total. In Scenario A, we fixed modulus $\rho_i^{(r)} = 1.001$, $i = 1, \cdots, 5,~r = 1, \cdots, 20,$ in Phase 1. We then changed the values of $\rho_2^{(r)}, \rho_5^{(r)}$ in Phase 2 and denoted $\delta(\rho)$ as the module difference between phases. In Scenario B, we allowed the modulus slowly increase by $5 \times 10^{-5}$ staring from $1.001$ in Phase 1 and various values in Phase 2. All the other parameters remained the same as previous simulation settings. Table~\ref{revision_permutationtest} summarizes the proportion of rejecting the null hypothesis based on $1500$ replications. It can be seen that the Type I error rates are close to the nominal level $.05$ and the power increases up to $1$ rapidly under both scenarios.

\begin{table} 
	\centering
	\caption{Type I error / power table of the proposed permutation test.}
	\label{revision_permutationtest}
	\begin{tabular}{ccccccc}
		\hline  \T \B 
		&		& $\delta(\rho) = 0$ & $\delta(\rho) = 1$ & $\delta(\rho) = 2$& $\delta(\rho) = 3$ &$\delta(\rho) = 4$    \T \B   \\  \hline
		
		\multirow{2}{*}{Scenario A}	&	$\rho_2$			& 0.056 & 0.110 & 0.586 &0.966  & 0.924  \T \B   \\ 
		&	$\rho_5$			&   0.046 & 0.112 &  0.600 & 0.992 &  0.955  \T \B  \\ \hline
		
		\multirow{2}{*}{Scenario B}	&		$\rho_2$			&  0.050 &  0.108 &  0.418 &  0.844 & 1.000  \T \B   \\
		&	$\rho_5$			&  0.047 &  0.268  &  0.586 & 1.000  & 1.000  \T \B   \\ \hline
	\end{tabular}
\end{table}

\subsection{Sensitivity analysis}
We have conducted extensive sensitivity analysis to investigate the performance of the proposed E-SSM when the underlying model assumption is violated, including when the number of AR(2) mixture components is mis-specified, and when the underlying singal deviates from AR(2) process. The discussions are presented in the Supplementary File, Section 4.

\section{Analysis of LFP data from olfaction sequence memory study}
\label{section_real}
\subsection{Data description}
The LFP dataset was obtained from an experiment searching for direct evidence of coding for the memory of sequential relationships among non-spatial events \citep{allen2016nonspatial}. During the course of the experiment, rats were provided with series of five odors multiple times. 
During the experiment, as rats performed the tasks, LFPs were recorded in the CA1 pyramidal layer of the dorsal hippocampus. 
The LFPs data set in this study comprise of 12 electrodes and
247 epochs. Each epoch is recorded over 1 second, aligned to port entry, sampled at 1000 Hertz and thus has $T = 1000$ time points. 

\subsection{Exploratory analysis}
We are interested in addressing two questions: (1) to determine how the original high-dimensional signals 
can be sufficiently represented by lower dimensional 
summary signals; and (2) to assess if and how the spectral properties 
of the LFP signals evolve across epochs during the experiment.  

To address the first question, we note the assertion in other 
studies (e.g., \cite{makarova2014can}) that the natural geometry 
of these neuronal assemblies gives rise to possible spatial 
segregation. This suggests that it is plausible to represent 
LFP data by lower dimensional summaries. 
In this nonspatial sequence memory study, we observe similar pattern across all the 12 electrodes. In Web Figure~21, although the power varies within each electrode, the synchrony of pattern across electrodes is still critical. For example, electrodes T13 and T14 behave almost identically. Electrodes T7, T8 and T9 also follow the same pattern during the course of experiment. 
Moreover, as part of this exploratory analysis, we implemented 
spectral principal component analysis \citep{brillinger1964frequency}, which is widely used in the exploratory analysis of brain imaging data \citep{wang2016exploratory}. 
Web Figure 7 (Supplementary File) presents the boxplots of the percentage of variability accounted by the first one and the first three components respectively. It can be shown that 3 components (mixture of delta, alpha and gamma bands) account for roughly 
$92\%$ of the variability with the first component accounting for 
$70\%$. All these findings validate the assumption that the original LFPs can be projected into low dimensional source signals without substantial loss of information. In this paper, we will build on this preliminary analyses by giving a more specific 
characterization of these signal summaries or components using the 
AR$(2)$ process. 

\begin{figure}[H]
	\includegraphics[width = .9 \textwidth, height = 0.3\textheight]{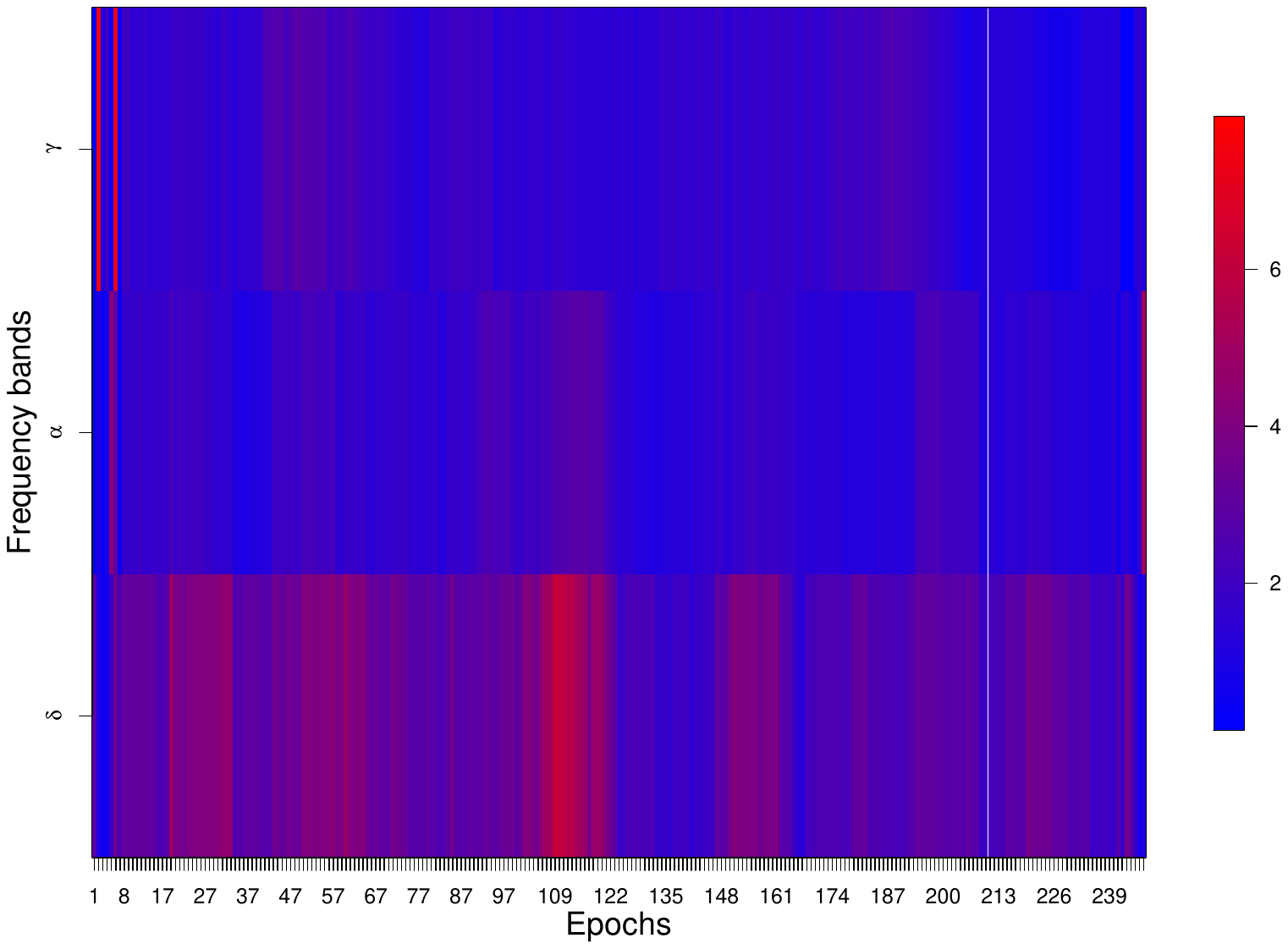} \vspace{-.1in}
	\caption{The evolution of power spectrum among delta (0-4 Hertz), alpha (8-12 Hertz) and gamma (30-35 Hertz) bands. Each band was averaged over all the electrodes.}
	\label{real_avg_tetrode}
\end{figure}

To gain insights into addressing the second question, we 
examined 
the log periodogram boxplots in Figure~\ref{Intro_ex2} across all the frequencies, 
we notice that the powers are quite spread out, especially at lower frequencies and the two peaks around delta and slow gamma bands. The heatmap in Figure~\ref{Intro_ex3} 
demonstrates the dynamics from early, middle, and late stages of the whole session. Web Figure~21 shows the evolving of the power across all the electrodes particularly on delta, alpha, and gamma bands. It shows that higher frequency bands dominate in early stage, while lower frequency bands capture more power during the evolution of experiment. In Figure~\ref{real_avg_tetrode}, an interesting pattern emerges: the burst of gamma activity on Phase 1 of the epochs is not replicated at other phases. One possible interpretation is that odor sequence (on which the animals have had extensive training) is re-encoded early in each session, which requires high frequency (gamma) activity, but later in the session, gamma activity is regulated and other lower frequencies (delta and alpha) become more prominent. 
Promoted by all these results, a further study is necessary to uncover the latent lower dimensional source signals that drive the observed LFPs. 

\subsection{Results and Discussion}
\begin{figure}  
	\begin{tabular}{c}
		\includegraphics[width = 1.1\textwidth, height = 0.3\textheight]{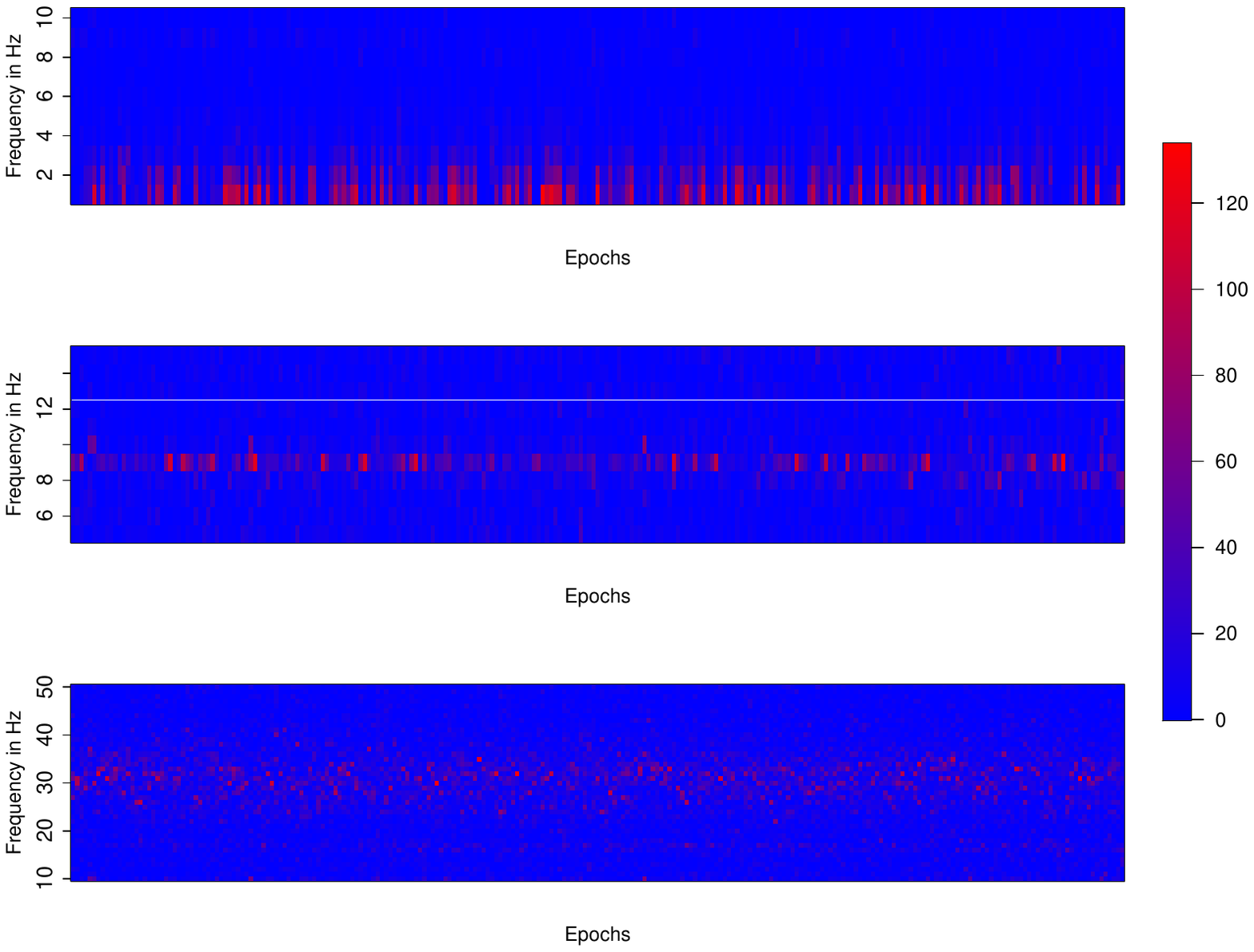} 
	\end{tabular}	 
	\caption{The periodograms of estimated latent  AR(2) processes corresponding to delta (top), alpha (middle) and gamma (bottom) bands.}
	\label{real_AR_heatmap}
\end{figure}

We applied our proposed E-SSM method to this study. 
Web Figure 8 (Supplementary File) shows time series plots of modulus (root magnitudes) corresponding to each of the three frequency bands as epochs evolve. In this plot, we could clearly identify the evolution of each individual module and a strong temporal dependence. Figure~\ref{real_AR_heatmap} displays the power of three latent source signals evolving during the period of experiment. We observe that the delta band captures the most power among all bands and is persistent across all phases. The alpha band attains its maximum power during the early phase and diminishes quickly in the middle stage and obtains more power in the end. There appear to be discontinuities in the delta, alpha and gamma power across the entire experiment. 
One interpretation to these results from the E-SSM analysis is that these on-off patterns could be just random variation. Another is that these are actual resetting of neuronal responses. This phenomenon of phase resetting in neurons is also observed in many biological oscillators. In fact, it is believed that phase resetting 
plays a role in promoting neural synchrony in various brain pathways. In either case, it is imperative 
to be cautious about blindly assuming that the neuronal process behaves identically across epochs. 
Doing so could produce misleading results. 

\begin{figure}[H]
	\includegraphics[width = .7\textwidth, height = 0.35\textheight]{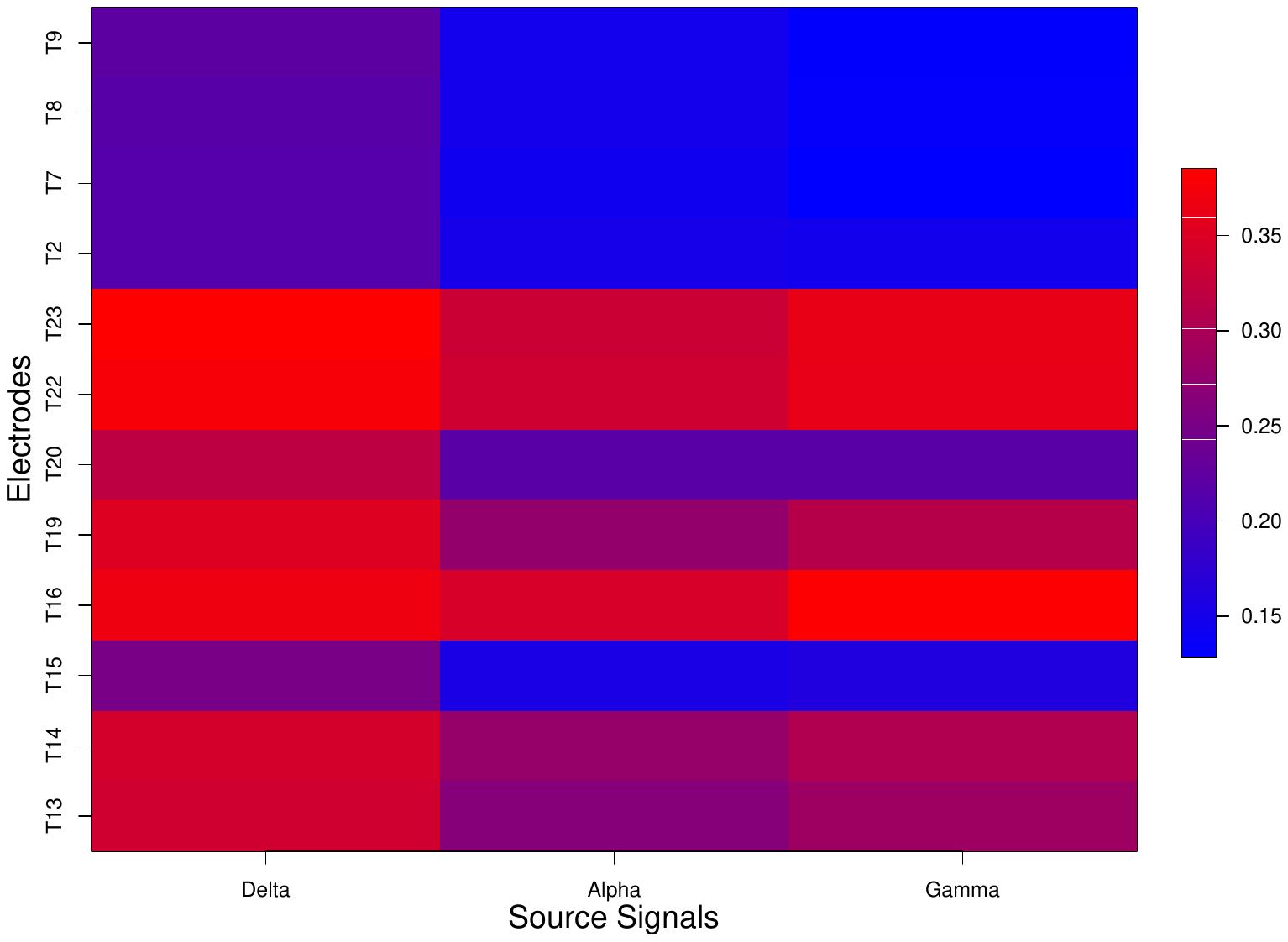}
	\caption{The estimated mixing matrix. Darker color represents heavier weights given by the latent processes (delta, alpha, gamma) on the LFPs.}
	\label{real_est_matrix}
\end{figure}

\begin{figure} 
	\centering
	\begin{tabular}{ccc}
		\includegraphics[width =0.3 \textwidth, height = 0.15\textheight]{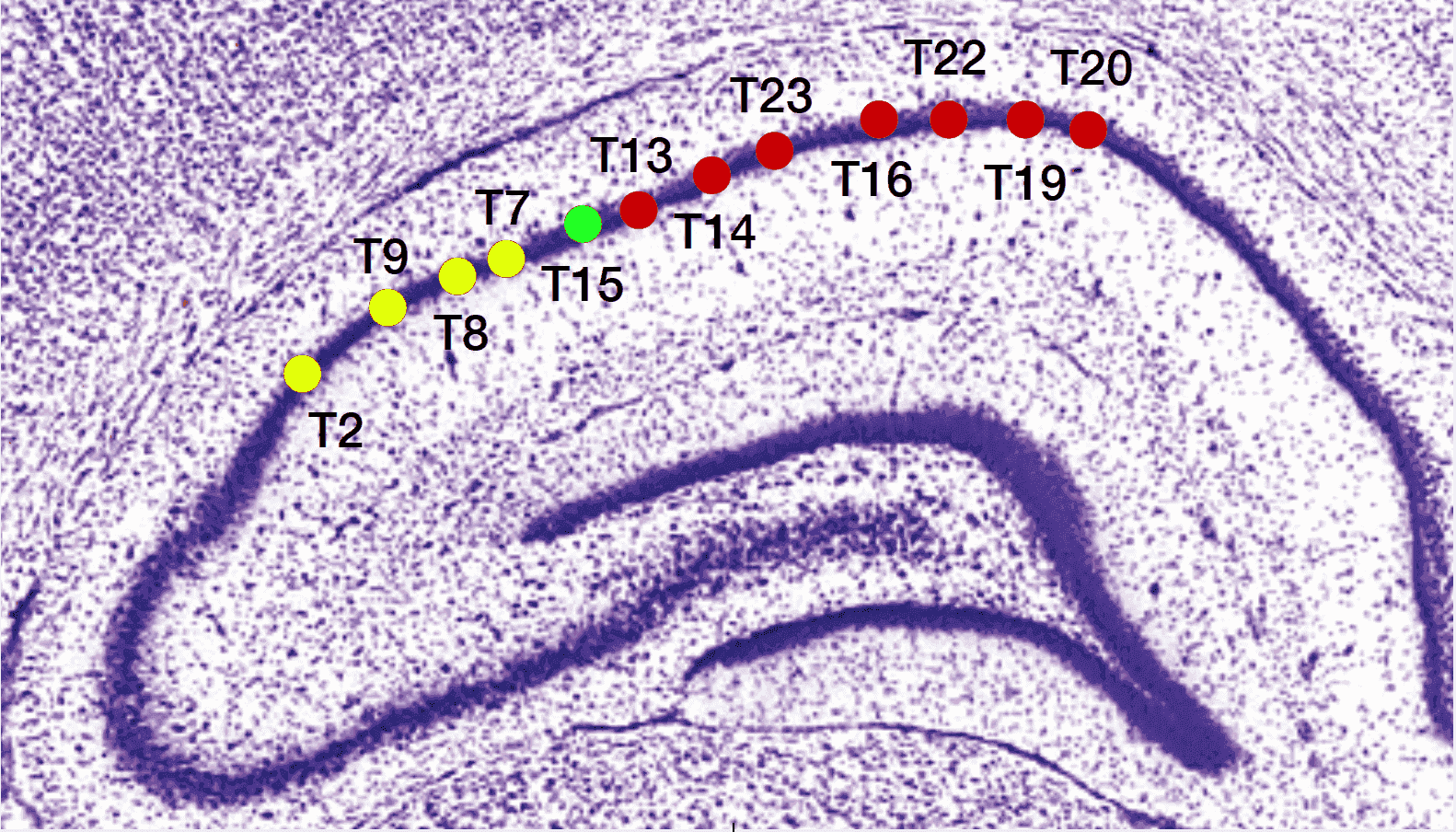}&	
		\includegraphics[width=0.3\textwidth,height=0.15\textheight]{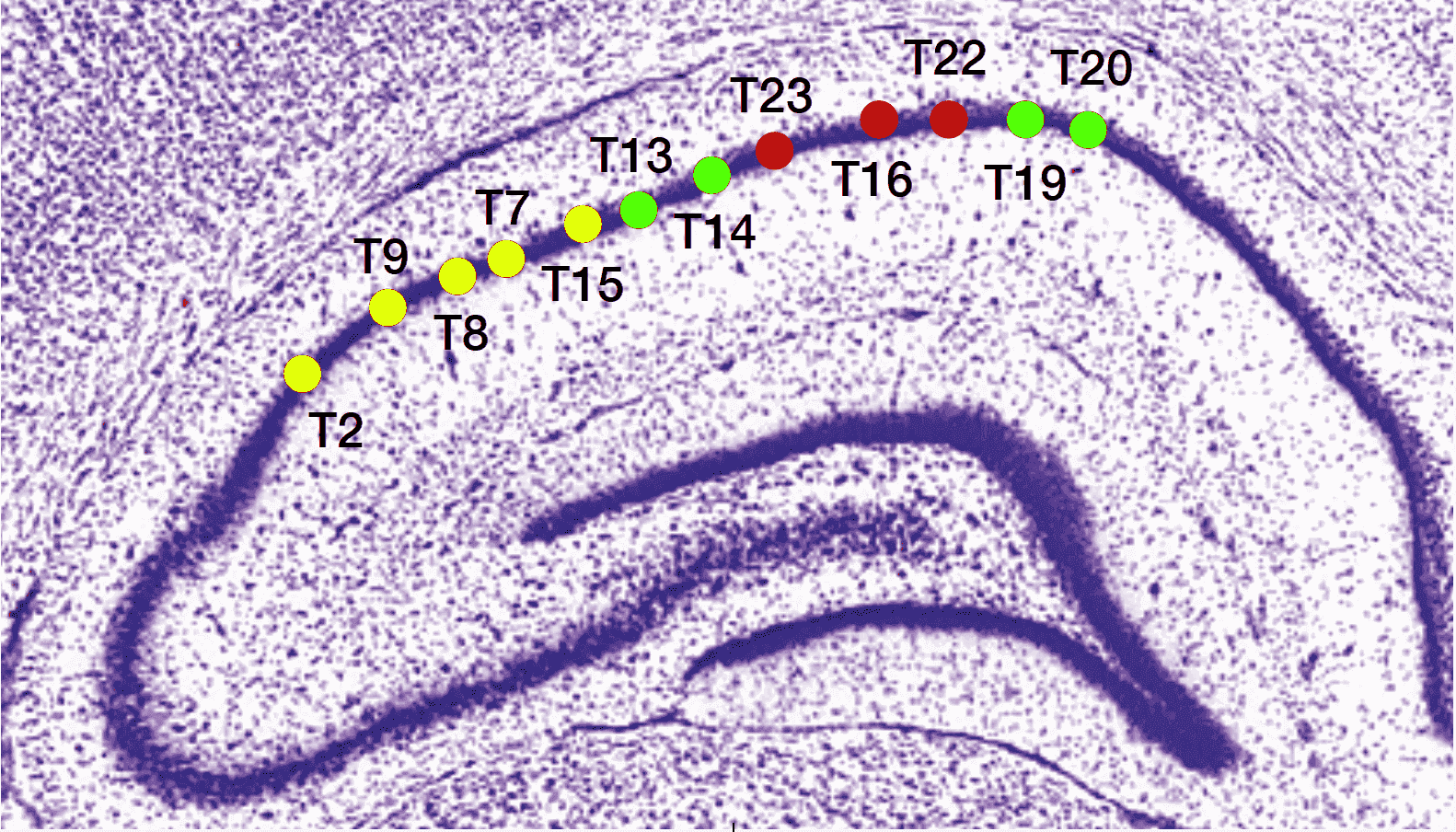}&
		\includegraphics[width=0.3\textwidth,height=0.15\textheight]{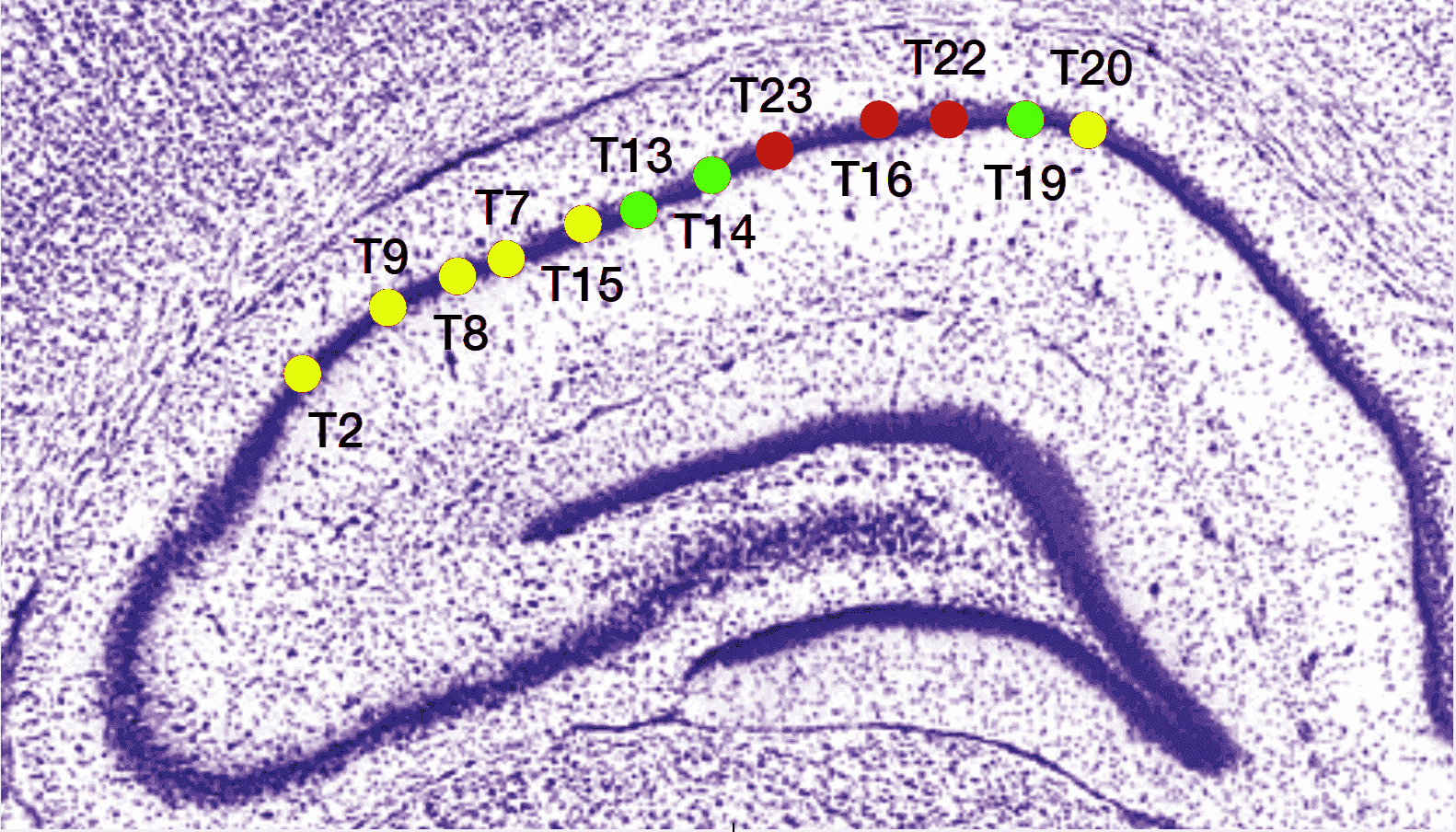}\\
		delta frequency band & alpha frequency band & gamma frequency band
	\end{tabular}
	\caption{Cluster analysis results among all the three frequency bands. Same color indicates the same cluster.}
	\label{real_cluster_mixing}
\end{figure}

We also study the mixing matrix to investigate how electrodes are associated across the three frequency bands. From Figure~\ref{real_est_matrix}, at delta band, electrodes T13, T14, T16, T19, T22, T23 are likely to be linked in terms of large power. Electrodes T15, T2, T7, T8 and T9 share the lowest power. At the alpha band, electrodes T16, T22 and T23 maintain the most power in contrast with electrodes T15, T2, T7-9 that obtain the lowest power. This pattern of association may result from the anatomical connections. Similarly, at gamma band, electrodes are connected in the same way as alpha band. We also used a cluster analysis on the entries of ``mixing'' matrix to understand the connection among electrodes. Similar to the results shown in Figure~\ref{real_est_matrix}, we are able to identify the same pattern in Figure~\ref{real_cluster_mixing}, through the visualization of cluster analysis. At delta band, electrodes T13, T14, T16, T19, T20, T22, T23 share the same pattern while T3, T7-9 are in the same cluster. Clusters at the alpha and gamma bands are roughly identical, which coincide with the results in Figure~\ref{real_est_matrix}. To the best of our knowledge, this approach (i.e., clustering of electrodes or 
nodes) has not be used previously for this kind of analysis. This has the potential for future explorations on synchrony among neuronal populations. Finally, we note here that the specific parametric AR(2) structure in our E-SSM has facilitated ease of 
interpretation of the oscillatory activity of these sources. 

Model validation and diagnostics were done using sample auto-correlations (ACF) and partial auto-correlations (PACF)  calculated from the residuals. 
Web Figure 9 (Supplementary File) shows an example of those values obtained from a representative electrode. We could easily observe the uncorrelated structure among the residuals. A p-value of $0.75$ based on the Ljung-Box test also provides some evidence to suggest white noise residuals and thus conclude that the proposed E-SSM 
fits this LFP data well.
 
\section{Concluding remarks}\label{conclude}
In this paper, we have proposed an evolutionary state space model (E-SSM) that allows the latent source signals to evolve across epochs. Although the results reported in this paper are quite promising, nevertheless, modeling the evolution/dynamics across epochs still remains a challenge in general. For example, we ignored the subject specific random effects in the current paper, which should be taken into account in a future work. 

In the simulation studies, we set the phase parameter to be the true values used in simulating the data. We also tried using the estimates from true data. This is done by first calculating the periodograms for each channel and trial, and then choosing the frequencies whose peaks were located as the phase parameters. Although these estimated phase parameters deviated slightly from the true ones in some cases, the estimated periodograms were able to capture the structure of the  true latent sources. These findings are consistent with the results from the sensitivity analysis.

In the paper, we choose to fix the location of spectrum peaks at pre-determined values. In frequency domain analysis, there are basically two approaches to obtain the power for particular bands: average and integral \citep{delorme2004eeglab}. To be more specific, average is more straightforward but neglects the range of frequency bands (e.g. theta: 4 - 8 Hertz vs lower beta: 12 - 18 Hertz). Integral is more complicated and more sensitive to the choice of range of frequency bands. A common observation in practice is that lower beta bands are usually more ``flatten-out" than theta bands when using average approach. The key point to both approaches is to find a ``center" for each frequency band. Back to our model, since each latent AR(2) corresponds to one particular band, we choose the center of range as the exact phase parameter for each frequency bands.  For example, we fix the phase at 10 Hertz for alpha band. There are a few reasons to do so: (1) From the existing literature, the power spectrum of particular frequency bands mostly achieve their peaks at the center within the range \citep{buzsaki2006rhythms}. (2) By fixing the peak beforehand, we can avoid identifiability issues. If we ``let the data drive the estimates of the location" as suggested in the comment, we could run into identifiability problem easily, i.e., we can change the columns of the mixing matrix and their corresponding AR(2) sources to get the same observed signals. (3) We have conducted some sensitivity analysis on different peaks within each particular bands. The results show that the ``constructed" signals are quite similar to the original ones. (4) It is reported from other recent studies (e.g., \cite{allen2016nonspatial}) that the approach of using centers of the range produces consistent and interpretable results. In the future research, it will be of interest to develop more flexible methodology that takes account the data uncertainty in determining the location of spectrum peaks. 

It is worth mentioning that the spectrum of a weakly stationary process being
approximated by the spectrum of an AR(2) mixture does not necessarily imply that the original process is approximated by the AR(2) mixture. The focus of this paper is motivated by the frequency domain analysis of the imaging data, where the actual LFP values are not as important as their frequency domain implications. It will be of interest in the future research to develop new models based on AR(2) mixture process (or any other meaningful basis in practice) within the time domain framework.

\section*{Acknowledgments}

Shen is supported by the National Science Foundation (DMS--1509023), and the Simons Foundation (Award 512620). Shahbaba is supported by NSF grant DMS1622490 and NIH grants R01MH115697 and R01AI107034. Fortin is supported by National Science Foundation (Awards IOS-1150292 and BCS-1439267) and
   Whitehall Foundation (Award 2010-05-84). The authors thank the Editor, an associate editor, and reviewers for helpful comments and suggestions.

\vskip 14pt
\noindent {\large\bf Supplementary File}

Technical proofs, additional simulation and data analysis results are provided in the supplementary file.  
\par
\par

\small
\bibliographystyle{chicago}
\bibliography{xg}


\end{document}